\documentclass[twocolumn,preprintnumbers,amsmath,amssymb]{revtex4}

\newtheorem{theorem}{Theorem}

\begin{document}


\title{Equivalence of Three-dimensional Spacetimes}

\author{F. C. Sousa\footnote{fcsousa@fisica.ufpb.br}, J. B. Fonseca\footnote{
jfonseca@fisica.ufpb.br}, C. Romero\footnote{cromero@fisca.ufpb.br}}

\affiliation{Departamento de F\'isica, Universidade Federal da Para\'iba \\
Caixa Postal 5008, 58059-900 Jo\~ao Pessoa - PB, Brazil.}

\begin{abstract}
A solution to the equivalence problem in three-dimensional
gravity is given and a practically useful method to obtain
a coordinate invariant description of local geometry is 
presented. The method is a nontrivial adaptation of Karlhede 
invariant classification of spacetimes of general relativity. 
The local geometry is completely determined by the curvature 
tensor and a finite number of its covariant derivatives in 
a frame where the components of the metric are constants. The results are
presented in the framework of real two-component spinors 
in three-dimensional spacetimes, where the algebraic 
classifications of the Ricci and Cotton-York spinors are given
and their isotropy groups and canonical forms are determined.
As an application we discuss G\"odel-type spacetimes in three-dimensional 
General Relativity. The conditions for local space and time 
homogeneity are derived and the equivalence of three-dimensional 
G\"odel-type spacetimes is studied and the results are compared 
with previous works on four-dimensional G\"odel-type spacetimes.

\end{abstract}

\maketitle

\section{Introduction}

The arbitrariness in the choice of coordinates is a basic
assumption underlying general relativity theory. 
This hypothesis gives rise to the equivalence problem, namely
the problem of deciding whether two spacetime metrics are 
different or are transformable one to another by a coordinate 
transformation. In other words, given two solutions of the 
field's equations, how to know whether they describe the 
same gravitational field? Furthermore, it can be difficult 
from a given metric to distinguish between real physical 
effects and those which depend only on the choice of coordinates. 
That is, the related question of how to decide whether certain effects 
have a physical origin or are due to the coordinate system used?
A solution of the equivalence problem provides a complete and invariant
characterization of the spacetime local geometry from which
the answers to these questions can be obtained.

From the mathematical point of view, the solution to the 
equivalence problem of $n$-dimensional Riemannian manifolds 
goes back to Christoffel~\cite{JE81} and the best approach 
was developed by Cartan~\cite{EC51}, which requires a 
comparison of curvature tensor components and their 
first $n(n+1)/2$ covariant derivatives. The development 
of computer algebra opened the way to the formulation of a 
procedure for testing equivalence of four-dimensional spacetimes 
in practice, that is, the Karlhede classification~\cite{K80,K06}. 
Finally, algorithms using both the Newman-Penrose spinor formalism 
and the algebraic classification of the irreducible parts of the 
curvature spinor were developed, enabling the implementation 
of the practical procedure in a computer algebra suite 
called {\scshape classi}~\cite{J87,MM83}, based on the computer 
algebra system for General Relativity {\scshape 
sheep}~\cite{Frick77,MM83,MM84,MM91,MS94}. For a review on 
the equivalence problem see~\cite{PSDI00a, PSDI00b, PSDI00c} 
and the references therein. It should be mentioned that, in order 
to deal with torsion, the equivalence problem techniques were 
generalized to Riemann-Cartan spacetimes~\cite{frt,frm} and 
implemented in a suite of computer algebra programs called {\scshape tclassi}~\cite{frm,afmr,frm1}, which is also based on {\scshape sheep}.

In this paper we present a solution to the equivalence 
problem of three-dimensional spacetimes and give a practical method
to obtain a coordinate invariant description of local geometry, 
which is presented by using spinor formalism. 
The method requires a nontrivial adaptation of Karlhede's 
invariant classification of spacetimes of general relativity. 

In the next section, we present a review Cartan's 
solution to the equivalence problem.
In section 3, Kalhede's invariant classification of 
spacetimes is presented. In section 4,we present
a brief review of the formalism of two-component
real spinors in three-dimensional spacetimes including
the curvature and the Cotton-York spinors, as well as
the Ricci and Bianchi identities. In section 5, we show
the algebraic classifications of the Ricci and Cotton-York
spinors, including their canonical forms and isotropy groups,
by using the spinor formalism. In section 6, we obtain 
a minimal set of components of the $n$-th derivatives 
of the Riemannian curvature spinor such 
that all derivatives of a given order $m$ can be expressed 
algebraically (using sums, products and contractions) 
in terms of these sets for $n\leq m$. This new result 
given here for three-dimensional spacetimes is  
analogous to the result obtained by MacCallum and 
{\AA}man~\cite{MA86} for four-dimensional spacetimes.
In section 7, three-dimensional G\"odel-type spacetimes are examined 
by using the equivalence problem techniques. The conditions for local space and time 
homogeneity are derived and the equivalence of three-dimensional 
G\"odel-type spacetimes is studied. An invariant classification is obtained
and the results are compared with previous works 
on four-dimensional G\"odel-type spacetimes. Finally,
in section 8 we present some conclusions.

\section{Cartan's Solution}

In this section we review Cartan's solution to the
equivalence problem, applied to pseudo-Riemannian
manifolds. The result is presented for $n$-dimensional
manifolds. This is a very general result, since it
is obtained for an arbitrary manifold without the use
of any gravitational field's equations.

The (local) gravitational field in general relativity is completely 
determined by the components $g_{\mu \nu}$ of the metric tensor in
a given coordinate system, which are a solution of 
Einstein's equations. Therefore, within the context of general 
relativity, (local) equivalence of spacetimes means (local) 
isometry of pseudo-Riemannian spacetimes.

In a more formal way, we say that two $n$-dimensional
spacetimes $M$ e $\widetilde {M}$, are (locally) equivalent
when there exists a diffeomorphism $f:U \mapsto \widetilde{U}$ 
between two coordinate systems $(U,x)$ and $(\widetilde{U},\tilde{x})$,
where $U\subset M$ and $\widetilde{U}\subset\tilde{M}$ are open sets 
defined on $M$ and $\widetilde{M}$, respectively, such that 
$\tilde{x} = f(x)$ and  
\begin{equation}
\tilde{g}_{\mu \nu} ( \tilde{x}) = \frac{\partial x^{\alpha} }
         {\partial \tilde{x}^{\mu} }
         \frac{\partial x^{\beta} }{\partial \tilde{x}^{\nu} } 
         g_{\alpha \beta}(x),
\label{trgeqvmet} %
\end{equation}
where $g_{\alpha\beta}(x)$ and $\tilde{g}_{\mu\nu}(\tilde{x})$
are the components of the metrics on $M$ and $\widetilde{M}$
with respect to the coordinate systems  $(U,x)$ and 
$(\widetilde{U},\tilde{x})$, respectively.

Despite the intuitive meaning of the equivalence definition
given by eq.\,(\ref{trgeqvmet}), its reformulation in terms 
of differential 1-forms is mostly desirable, taking into account
Cartan's method to determine the equivalence of 
sets of \mbox{1-forms}. 
Let $\omega^{a}=\omega^{a}_{\ \mu}dx^{\mu}$ 
and ${\tilde \omega}^{a}={\tilde \omega}^{a}_{\ \mu}d{\tilde x}^{\mu}$
be non-holonomic coframes uniquely defined in coordinates systems 
$(U, x)$ and $({\widetilde U}, {\tilde x})$ on $n$-dimensional 
manifolds $M$ and ${\widetilde M}$, respectively. We say that
these sets of linearly independent 1-forms are equivalent when
there exists a coordinate transformation $x^{\mu}=x^{\mu}({\tilde x})$ 
such that ${\tilde\omega}^{a}=\omega^{a}$. Cartan~\cite{EC51} 
showed that these non-holonomic coframes are equivalent
if, and only if, the system of algebraic equations obtained by
the comparison of the non-holonomic objects $C^a_{\ pq}$ 
and ${\tilde C}^a_{\ pq}$ and their covariant derivatives 
according to
\begin{eqnarray}
\widetilde{C}^{a}_{\ pq}(\tilde{x}) & = 
    & C^{a}_{\ pq}(x), \nonumber \\
\widetilde{C}^{a}_{\ pq;m_{1}}(\tilde{x}) & = 
    & C^{a}_{\ pq;m_{1}}(x), \nonumber \\
    & \vdots &  \label{teocart} \\
\widetilde{C}^{a}_{\ pq;m_{1} \ldots m_{(p+1)} }(\tilde{x}) & = 
    & C^{a}_{\ pq;m_{1}
      \ldots m_{(p+1)} }(x), \nonumber  
\end{eqnarray}
is compatible, that is, there exists a solution 
$x^{\mu}=x^{\mu}(\tilde{x})$. Here, and in what 
follows, the covariant derivative is denoted by a semi-colon.

The non-holonomic objects and their covariant derivatives at each member
of eqs.\,(\ref{teocart}) are obtained by calculating 
the exterior derivatives
\begin{eqnarray}
d\omega^{a} &=& 
  \frac{1}{2}C^{a}_{\ pq}\,\omega^{p}\!\wedge\omega^{q},
  \label{objnh} \\
dC^{a}_{\ pq} &=& 
  C^{a}_{\ pq;m}\,\omega^{m},  \label{d1objnh} \\
dC^{a}_{\ pq;m} &=& C^{a}_{\ pq;mn}\,\omega^{n}   
  \label{d2objnh} 
\end{eqnarray}
and successively. 
The covariant derivative of order $p+1$ is the 
lowest order derivative which is functionally dependent
on the elements of the set $\{ C^{a}_{\ pq}, C^{a}_{\ pq;m_{1}}, 
\cdots,  C^{a}_{\ pq;m_{1}\ldots m_{p}} \}$, given by
all lower derivatives up to order $p$. That is,
the $p+1$ derivative is expressible in terms of its 
predecessors. Since each derivative either gives 
(at least) one new functionally independent 
function or is the last we need to consider, we
obtain the limit $(p+1) \leq n$, taking into account
that there exist at most $n$ functionally independent
functions on an $n$-dimensional manifold. 

The most appropriate context to deal with the
equivalence problem according to Cartan's method
is the bundle $F(M)$ of generalized orthogonal frames
defined over a pseudo-Riemannian manifold $M$. The
frame bundle $F(M)$ is a differentiable manifold
whose points are given by a pair: a point $p$ of
$M$ and a generalized orthogonal frame defined on
$p$. That is, a manifold given by
$F(M)=\bigcup_{p\in M}F_{p}$, where $F_{p}$ is the
set of all generalized orthogonal frames defined
at $p\in M$, called the fiber over $p$. 
These frames are given by the linearly
independent vector fields $h_{a} = h_{a}^{\ \mu}(x)\partial_{\mu}$
($a=1,\ldots,n$), where the components of the metric $\eta_{ab}=
g(h_{a},h_{b}) = g_{\mu \nu}h_{a}^{\ \mu} h_{b}^{\ \nu}$ are 
constant and given by a symmetric matrix $\eta=(\eta_{ab})$, 
with the appropriate signature~\cite{MM83}. 

The generalized orthogonal frames cannot be used
to define equivalence, since they are not uniquely
defined. There exist linear transformations 
$h_a\mapsto\Lambda_a^{\ b}h_b$ 
which leave invariant the components of the metric 
$\eta_{ab} = \Lambda_{a}^{\ c}\,\eta_{cd}\,\Lambda_{b}^{\ d}\,$. 
These transformations are called generalized rotations and
form the group $O(n)$ with $\frac{1}{2}n(n-1)$ 
parameters~\cite{MM83, KSMH80, HE73}.

Underlying Cartan's approach to solve the equivalence
problem there is the fact that equivalent $n$-dimensional
pseudo-Riemannian spacetimes $M$ and $\tilde{M}$ 
have equivalent bundles of generalized orthogonal frames $F(M)$ 
and $F(\tilde{M})$, respectively. Locally, $F(M)$ is the 
product $U \times O(n)$ of the subset $U\in M$ and the generalized
orthogonal group $O(n)$. The fiber $F_p$ over a point $p$ of $U$ is 
isomorphic to the generalized orthogonal group $O(n)$, since it is the
set of all generalized orthogonal frames defined at $p$, which 
are related by generalized orthogonal transformations. Thus, the
coordinates of $F(M)$ are given by the $n$ coordinates $x=(x^a)$ of
the point $p$ and the $n(n+1)/2$ coordinates (parameters)
$\xi=(\xi^A)$ of the generalized orthogonal group $O(n)$.

The crucial point for Cartan's approach is the requirement
of a set of uniquely defined linearly independent 1-forms. 
Since the freedom in the choice of generalized orthogonal frames
in $M$ is lost in $F(M)$, there is a uniquely-defined basis of
the cotangent space $T^\ast_{\,P}(F(M))$, given by both the
canonical 1-form $\Theta^A=H^A_{\,\,\mu}(x,\xi)dx^{\mu}$ and the connection 1-form $\Sigma^A_{\,\,B}=\Gamma^A_{\,\,B\mu}(x,\xi)dx^{\mu} 
+ \Gamma^A_{\,\,B\xi}(x,\xi)d\xi^{A}$
of $F(M)$. Therefore, we can reformulate the definition of
(local) equivalence given by eq.\,(\ref{trgeqvmet}) in the
following way~\cite{K06,JE81,MM83,KobNom63,Stenbe64,Spivak79}.
Let $M$ and ${\widetilde M}$ be two $n$-dimensional
pseudo-Riemannian manifolds and $F(M)$ 
e ${\widetilde F}({\widetilde M})$ the fiber bundles
of generalized orthogonal frames over $M$ and 
${\widetilde M}$, respectively. We say that $M$ and 
${\widetilde M}$ are (locally) equivalent when there
exists a local diffeomorphism  
$J:F(M)\mapsto {\widetilde F}({\widetilde M})$ such that
\begin{equation}
J^\ast \widetilde{\Theta}^A =\Theta^A \;\;\;\mbox{and}\;\;\; 
J^\ast\widetilde{\Sigma}^A_{\,\,B} = \Sigma^A_{\,\,B}
\end{equation}
hold. Here $J^\ast$ is the pull-back map defined from $J$.

A solution to the equivalence problem for 
pseudo-Riemannian manifolds can then be obtained by using
Cartan's result on the equivalence of sets of 1-forms
together with Cartan's equations of structure for a 
pseudo-Riemannian manifold. The solution can be summarized
as follows~\cite{K06,JE81,MM83,KobNom63,Stenbe64,Spivak79}. 
Two n-dimensional pseudo-Riemann manifolds
$M$ and $\tilde{M}$ are locally equivalent if and only if
there exists a local diffeomorphism $J$ between their corresponding
generalized orthogonal frame bundles $F(M)$ and $F(\tilde{M})$, such that the
following system of algebraic equations relating the components of
the curvature tensor and its covariant derivatives
\begin{eqnarray}
R^{A}_{\ BCD} & = &  \widetilde{R}^{A}_{\ BCD}\;, \nonumber \\
R^{A}_{\ BCD;M_{1}} & = & \widetilde{R}^{A}_{\ BCD;M_{1}}\;, \nonumber  \\
                  & \vdots &   \label{eqvcond} \\
R^{A}_{\ BCD;M_{1}\ldots M_{p+1}} & = & \widetilde{R}^{A}_{\ BCD;M_{1}                                             \ldots M_{p+1}}\;,\nonumber 
\end{eqnarray} 
are compatible as equations in generalized orthogonal frame 
bundle coordinates $\left( x^{a}, \xi^{A} \right)$.
The $(p+1)$-th derivative of curvature is the lowest 
derivatives which is functionally dependent on all the previous derivatives. 
The number of functionally independent functions is at most the dimension
of $F(M)$. Since there is at least one new functionally 
independent function at each order of derivative, it
follows that $p+1\leq n(n+1)/2$. 

In association with the necessary and sufficient
conditions for (local) equivalence given by
eqs.\,(\ref{eqvcond}), Cartan's solution also 
shows that all (local) metric properties of an arbitrary 
$n$-dimensional pseudo\--Riemannian manifold are 
described in a comprehensive and unique way by the set
\begin{equation}
I_{p} = \{ R^{A}_{\ BCD}\,,  
 R^{A}_{\ BCD;M_{1}}\,,\ldots,  
 R^{A}_{\ BCD;M_{1}\ldots M_{p}} \},  
\label{invcar}
\end{equation}
whose elements are called Cartan's invariants, since they 
are invariant under coordinate transformations on the base manifold.
But they depend on the orientation of the frame and change under
generalized orthogonal rotations. The theoretical upper bound for the number
of covariant derivatives to be calculated is $n(n + 1)/2$.

In principle, we can use the set $I_p$ of Cartan's invariants
to obtain all (local) properties of a spacetime  
that can be obtained from the components $g_{\mu\nu}$ of the 
metric in a given coordinate system. There are several
results where the Cartan's invariants are used to investigate
(local) properties of spacetimes. In the context of General
Relativity we have, for instance, the determination of the 
spacetime isometry group~\cite{KarMac82, ArDrSk92}, 
the investigation of limits of families of 
spacetimes~\cite{PaReMa93} and the local degrees 
of freedom on a spacetime~\cite{MM06}.

Concluding this section we review the results where
the dimensions of the isometry group and its isotropy 
subgroup are obtained from Cartan's invariants.

We say that a vector field with components $v^{\alpha}$ 
in a given coordinate system, defines a local isometry on
a pseudo-Riemannian manifold $M$ if, and only if, the
following conditions are satisfied:
\begin{equation}
\pounds_{v}g_{\mu \nu} = v_{\mu ;\nu} + v_{\nu ;\mu} = 0\,.
 \label{kileq} 
\end{equation}
The eq.\,(\ref{kileq}) above are called Killing's equations
and their solutions are Killing vector fields. They are
the generators of the isometry group and the maximal 
number of linearly independent Killing vectors is the 
dimension of the isometry group. There is an isotropy subgroup
when there exist Killing vector fields which generate 
one-parameter groups of transformations which leave invariant 
the points of the spacetime manifold. They are the generators 
of the isotropy subgroup.

According to Cartan's solution to the equivalence
problem, we can say that for each (local) isometry
of a manifold $M$ there exits, in bijective 
correspondence, a diffeomorphism on the bundle of 
generalized orthogonal frames $F(M)$ which
preserve the set $I_p$ of Cartan's invariants. Therefore, when
$I_p$ has $k_p$ functionally independent Cartan's invariants, 
then the system of algebraic equations eqs.\,(\ref{eqvcond}) 
has $k_p$ linearly independent equations and its solution
is a diffeomorphism on the frame bundle $F(M)$ which
depends on $\frac{1}{2}n(n+1)-k_p$ arbitrary constants. 
Dealing separately with the $t_p$ functions of the spacetime
coordinates $(x^{\mu})$  and the $m_p=k_p-t_p$ functions of the parameters
$(\xi^{A})$ of the generalized orthogonal group $O(n)$, it
can be shown that~\cite{EC51,K80} there
exists an isometry group of dimension $r$, with an 
isotropy subgroup of dimension $s$ and acting on a orbit 
of dimension $d$, where
\begin{eqnarray} 
s &=& \frac{1}{2}n(n-1)-m_{p}\,,
\label{idim}  \\
r &=& \frac{1}{2}n(n+1)-k_{p}=s+n-t_p\,,
\label{gdim}  \\
d &=& r-s=n-t_p\,.
\label{ddim}
\end{eqnarray}

\section{Karlhede Classification}

In this section we show how the difficulties to
deal with Cartan's solution of the equivalence
problem in practice are considerably reduced by
a procedure to test equivalence, based
on an algorithm developed by Karlhede, where all 
calculations are performed on the spacetime base 
manifold and the maximal order of the derivatives 
is reduced~\cite{K80}. 

Cartan's solution of the equivalence problem has
an important aspect to be considered in practice.
At each order $q=0,1,\ldots,(p+1)$ of derivative 
of the curvature tensor, there are several 
properties which can be determined. Their comparison
constitute necessary conditions to test the equivalence
and can be used at each order $q$ to establish 
a practical procedure. The test finishes whenever
one of these conditions is not satisfied. 
Only when all necessary conditions for all orders
$q=0,1,\ldots,(p+1)$ of derivatives are satisfied
it is necessary to verify the consistency of the system 
of algebraic equations given by eqs.\,(\ref{eqvcond}), 
that is, the necessary and sufficient conditions
for equivalence. 
The relevance of this approach is evident from
the practical point of view when considering that
there is no procure which makes formally decidable
the problem of verifying whether or not a system of 
algebraic equations has a solution~\cite{MS94}. 

Following this approach, we will describe 
the steps needed in order to present Karlhede's 
algorithm to test the equivalence.
Initially it is necessary to handle separately the spacetime
manifold coordinates $x^{\mu}$ and the parameters $\xi^{A}$ of
the generalized orthogonal group. This is done by 
calculating the Cartan's invariants in a section of
the bundle of generalized orthogonal frames, that is,
with respect to a given generalized orthogonal frame.
Therefore, all Cartan's invariants are calculated on the
base spacetime manifold and no more depend on the parameters
of the group of generalized rotations. Thus, now the set
\begin{equation}
I_p = \{ R^{a}_{\ bcd},\,\ldots,\; 
             R^{a}_{\ bcd;m_{1} \ldots m_{(p+1)}}\}  
\label{ipdef}
\end{equation}
is given by the components $R^{a}_{\ bcd}$ of the curvature
on $M$ and their covariant derivatives, with respect to
the generalized orthogonal frame which define the local
section of the frame bundle.
Nevertheless, the dependence of Cartan's invariants on 
the parameters $\xi^{A}$ of the generalized orthogonal group
still can be verified through their behavior under generalized
rotations.

The next step deals with the reduction, at
each order of differentiation $q=0,1,\ldots,(p+1)$, of the
dimension of the fiber bundle effectively used to test
the equivalence. This is achieved by choosing a 
generalized orthogonal frame which is aligned with invariant
directions determined by Cartan's invariants. With this 
choice the freedom of generalized rotations is reduced and the frame
is fixed as much as possible. These invariant directions are
determined by the algebraic classifications of the Cartan's
invariants, which define canonical forms for each one of them. 
These generalized orthogonal frames are called standard or
canonical frames~\cite{K80}.
This step has two important aspects from a practical point
of view. First, along with the reduction of the dimension of the effective
frame bundle the maximal order of derivatives is also reduced.
Second, the algebraic classifications define a set of necessary
conditions given by the algebraic types, their canonical
forms and groups of isotropy.

The reduction of the dimension of the effective frame
bundle at each order $q=0,1,\ldots,(p+1)$ of derivatives,
can be determined from the isotropy group $H_{q}$ of the set $I_{q}$. 
This is the group of generalized rotations which leave
invariant the canonical forms of the elements of $I_{q}$.
Thus, when two spacetimes have the same isotropy
group $H_{q}$, the number of functionally independent
functions of the parameters $\xi^{A}$ are the same
as well~\cite{MM83}. Therefore, the freedom of generalized
rotations of the canonical frame is reduced each time
the dimension of $H_{q}$ is less
than the dimension of $H_{(q-1)}$. Since $H_{q}$ is
a subgroup of $H_{(q-1)}$, the parameters of the
generalized rotations which do not belong
to $H_{q}$ can be used to fix furthermore the canonical frame.

The next step take into account the fact that local 
equivalence also requires that the Cartan's
invariants have the same dependence on the coordinates
$x^{\mu}$ of the spacetime manifold. Thus, at each order
$q=0,1,\ldots,(p+1)$ of derivative it is necessary that
equivalent spacetime manifolds have the same
numbers $t_{q}$ of functionally independent functions
of the coordinates $x^{\mu}$ in the elements of $I_{q}$.

Finally, according to Cartan's solution of the
problem of equivalence, the Cartan's invariants
must be calculated until a order $q=(p+1)$ 
of derivatives where the elements of $I_{(p+1)}$ 
are functionally dependent on the elements 
of $I_{p}$.  Therefore, the steps of the practical
procedure finish at an order $q$ of derivative
where not only the groups $H_{(q+1)}$ and $H_{q}$
are the same, but also the numbers $t_{(p+1)}$
and $t_p$. The last step is to check the 
occurrence of these equalities.

All the steps of the practical procedure discussed
above can be joined in an algorithm which
starts by setting $q = 0$ and has the following steps \cite{K80}:
\begin{enumerate}
\item 
Calculate the set $I_{q}$, i.e.,  
the derivatives of the curvature up to the $q$-th order.
\item 
Fix the frame, as much as possible, by putting
the elements of $I_{q}$ into canonical forms.
\item 
Find the frame freedom given by the 
isotropy group $H_{q}$ of transformations which
leave invariant the canonical forms of $I_q$.
\item 
Find the number $t_{q}$ of functionally 
independent functions of spacetime coordinates 
in the elements of $I_q$, brought into the canonical 
forms.
\item   
If the isotropy group $H_{q}$ is the same
as  $H_{(q-1)}$ and the number of functionally independent
functions $t_{q}$ is equal to $t_{(q-1)}$,
then let $q=p+1$ and stop. Otherwise, increment 
$q$ by 1 and go to step $1$.    
\end{enumerate}
This procedure provides a discrete characterization of 
n-dimensional pseudo-Riemannian spacetimes, called
Karlhede classification, in terms of 
the following properties: the set of canonical forms in $I_{p}$, 
the isotropy groups $\{H_{0},\ldots ,H_{p}\}$ and the 
number of independent functions $\{t_{0}, \dots ,t_{p}\}$. 

To check the equivalence of two pseudo-Riemannian spacetimes
the above discrete properties of their Karlhede's classifications
are compared and only when they match is it necessary to 
determine the compatibility of eqs.(\ref{eqvcond}).

An important result obtained with the application
of Karlhede algorithm is the reduction of the maximal order
of derivatives, given by $n(n+1)/2$ according to
Cartan's theorem.

In the context of General Relativity, we
have four-dimensional pseudo-Riemannian spacetimes.
The canonical frame is fixed through the principal
directions of the Weyl tensor, obtained by the
Petrov's classification. Since the isotropy groups
os the Petrov types (except type 0) have dimension
$s\leq 2$, we obtain the limit $(p+1)\leq 7$. For conformally
flat spacetimes, the Weyl tensor vanish 
(Petrov type 0) and the principal directions of the
Ricci tensor are used instead. As the isotropy
groups of the Segre types (except Segre 0) of the Ricci tensor have
dimensions $s\leq 3$, we obtain the limit $p+1\leq 8$~\cite{MS94}.

For three-dimensional spacetimes, the Weyl tensor
vanishes identically and the canonical frame is
aligned with the principal directions of the Ricci
tensor. Now the Segre types (except type 0) have
isotropy groups of dimensions $s\leq 1$, and
we find the limit $p+1\leq 5$. 

The Karlhede algorithm was implemented by using
spinor formalism to deal with equivalence in General
Relativity, since it enables symmetries which are
complicated in tensorial form to be expressed in a
simple way. Since the same simplifications occurs
with respect to spinors in three-dimensional spacetimes,
in the next section we briefly present the two-component real
spinors in three-dimensional spacetimes and obtain
some results required to implement the Karlhede
algorithm.

\section{Two-component real spinors}

We shall consider three-dimensional spacetimes described by a metric 
with signature ($- + +$). In this section we present some results 
obtained by using the formalism of two-component {\em real} 
spinors~\cite{TC94,AN95,TG03}, which is analogous to the
Newman-Penrose~\cite{PR84a,PR84b,KSMH80} formalism of 
two-component complex spinors in four-dimensional spacetimes. It
should be mention that only one type of spinor index is
required for three-dimensional space. The spin transformation
are the elements of the group $SU(1,1)$ for two-component complex spinors
and the group $SL(2,R)$ for two-component real spinors~\cite{TC94,TG03}.

An one-index spinor will be denoted by $\psi^A$ or $\psi_A$ and
has two real components, where capital latin indices take
the values $0, 1$. These indices will be raised and lowered
by the Levi-Civita symbol $\epsilon_{AB} =-\epsilon_{BA}$ 
( $\epsilon_{01} = 1$), according to
\begin{equation}
\psi_A =\psi^B\epsilon_{BA}\,,\;\;\psi^A=\epsilon^{AB}\psi_B\,.
\end{equation}
The inner product is given by $\psi^A\phi_A=\epsilon^{AB}\psi_B\phi_A$ 
and the spin frame $\{o^A, \iota^A\}$ is normalized by $o_A\iota^A = 1$,
where $o^A=(1,0)$ and $\iota^A=(0,1)$.

The 2-order symmetric real spinors $\phi_{AB}$ corresponds
to vectors in three-dimensional spacetime $M$. There is a correspondence 
between a null frame of real vectors
$\{k^a, m^a, n^a\}$ ( or a Lorentz frame $\{t^a, x^a, z^a\}$ ) in
$M$ and a spin frame $\{o^A,\iota^A\}$ given through the real and
symmetric connecting quantities $\sigma^a_{\; AB}$ according to
\begin{eqnarray}
k^a &=& (t^a + z^a)/\sqrt{2}=\sigma^a_{\; AB}o^A\,o^B=\sigma^a_{\; 00}\,, \nonumber \\
n^a &=& (t^a - z^a)/\sqrt{2}=\sigma^a_{\; AB}\iota^A\iota^B=\sigma^a_{\; 11}\,,
\label{bases} \\
m^a &=& x^a =\sigma^a_{\; AB}(o^A\iota^B + o^B\iota^A)/\sqrt{2} = \sqrt{2}\,\sigma^a_{01}\,, \nonumber
\end{eqnarray}
where $m^a = x^a$ and $z^a$ are space-like, $t^a$ is time-like,
$k^a$ and $n^a$ are light-like. The correspondence between
the spaces also requires the correspondence between their inner
products
\begin{eqnarray}
g_{ab}   &=& \sigma_a^{\; AB}\sigma_b^{\; CD}g_{ABCD} \nonumber  \\ 
         &=& -k_a\,n_b-k_b\,n_a+m_a\,m_b  \nonumber \\
         &=& -t_a\,t_b + z_a\,z_b + x_a\,x_b\,,  \label{stmet} \\
g_{ABCD} &=& \sigma^a_{\; AB}\,\sigma^b_{\; CD}\,g_{ab} \nonumber \\
         &=& -\frac{1}{2}(\epsilon_{AC}\,\epsilon_{BD} 
             + \epsilon_{AD}\,\epsilon_{BC}). \label{spmet}
\end{eqnarray}

Now we can show that, corresponding to the normalization
$o_A\iota^A = 1$ of the spin frame, we have
\begin{equation}
k_an^a = t_at^a = -1 , \;\; m_am^a = x_ax^a = z_az^a = 1, 
\label{frcontr}
\end{equation}
with all other contractions vanishing identically. In general,
an arbitrary vector $v^a$ in the null frame eqs.\,(\ref{bases}) above
corresponds to a symmetric real spinor $\phi_{AB}$ according to
\begin{equation}
v^a = \sigma^a_{\; AB}\phi^{AB}\,,\;\; \phi^{AB} =-\sigma_a^{\; AB}v^a.
\end{equation}
Now, taking into account eqs.\,(\ref{stmet})-(\ref{spmet}), we obtain 
\begin{equation}
v_av^a = g_{ab}v^bv^a=g_{ABCD}\phi^{CD}\phi^{AB}=-\phi_{AB}\phi^{AB}.
\end{equation}
The spinor $\epsilon_{AB}$ fulfills the identity
\begin{equation}
\epsilon_{A[B}\,\epsilon_{CD]} = 0 , \label{jacobi}
\end{equation}
where square parentheses denote skew symmetrization.
One consequence of the identity eq.\,(\ref{jacobi}) above is the identity
\begin{equation}
\phi_A\psi_B - \phi_B \psi_A = \epsilon_{AB}\phi_N\psi^N, 
\label{asym}
\end{equation}
for any spinors $\phi_A$ and  $\psi_A$, which will be used repeatedly
to obtain the decomposition into irreducible parts of the
curvature spinor.

For three-dimensional spacetimes the Weyl tensor vanishes
identically and the Riemannian curvature tensor
can be decomposed in terms of the Ricci tensor 
$R_{ab}=R^c_{\,\,acb}$ and the scalar of curvature $R$. Using the 
traceless Ricci tensor $S_{ab} = R_{ab} -\frac{1}{3}g_{ab}R$,
the Riemann tensor is given by~\cite{Eisen}
\begin{eqnarray}
R_{abcd} &=& g_{ac}S_{bd} - g_{ad}S_{bc} + g_{bd}S_{ac} - g_{bc}S_{ad} \nonumber \\
         && -\frac{1}{6}(g_{ad} g_{bc} - g_{ac} g_{bd})R. 
\end{eqnarray}
and we can calculate the curvature spinor which, by using
the identity eq.\,(\ref{asym}), is decomposed into irreducible parts
in accordance with
\begin{eqnarray}
R_{AXBYCZDW} &=& \epsilon_{XY} R_{ABCZDW} \nonumber \\
             && +\epsilon_{AB}R_{XYCZDW} , \\
R_{ABCZDW}   &=& \epsilon_{ZW}Q_{ABCD} \nonumber \\
             && + \epsilon_{CD}Q_{ABZW} , \\
Q_{ABCD}     &=&  \Phi_{ABCD} - \frac{1}{3}g_{ABCD}\Lambda. 
\end{eqnarray}
Therefore, the irreducible parts of the three-dimensional curvature
spinor are the totally symmetric spinor $S_{ABCD}=2\Phi_{ABCD}$,
which corresponds to the traceless Ricci tensor $S_{ab}$, and
the curvature scalar $\Lambda = R$. For the Ricci tensor $R_{ab}$ we
obtain the spinor
\begin{equation}
R_{ABCD} = 2\,\Phi_{ABCD} + \frac{1}{3}g_{ABCD}\Lambda. 
\end{equation}
Similar results have been obtained in \cite{AN95}, but 
with a different signature of the metric $(+--)$ and a different
choice of $\Lambda$ in order to resemble the results for four-dimensional 
spacetimes. 

Now we can express the Ricci spinor $\Phi_{ABCD}$ using the
following abbreviations for the null frame components
of the traceless Ricci tensor $S_{ab}$. We define a real
symmetric $\Phi_{AB}$ ($A,B=0,1,2$) by
\begin{eqnarray}
\Phi_{00} &:=& \Phi_{0000} = \frac{1}{2}S_{ab}k^ak^b = \frac{1}{2}R_{ab}k^ak^b,
\label{phi0} \\
\Phi_{22} &:=& \Phi_{1111} = \frac{1}{2}S_{ab}n^an^b = \frac{1}{2}R_{ab}n^an^b, \\
\Phi_{10} &:=& \Phi_{1000} =\frac{1}{\sqrt{2}}(\frac{1}{2}S_{ab}m^ak^b) \nonumber \\
          &=& \frac{1}{\sqrt{2}}(\frac{1}{2}R_{ab}m^ak^b), \\
\Phi_{12} &:=& \Phi_{1011} =\frac{1}{\sqrt{2}}(\frac{1}{2}S_{ab}m^an^b) \nonumber \\
          &=& \frac{1}{\sqrt{2}}(\frac{1}{2}R_{ab}m^an^b), \\
\Phi_{11} &:=& \Phi_{0011} =\frac{1}{2}S_{ab}k^an^b \nonumber \\
          &=& \frac{1}{6}(R_{ab}m^am^b + R_{ab}n^ak^b). \label{phi2} 
\end{eqnarray}
Note that $\Phi_{AB}$ has only five independent components,
since we have the identity $\Phi_{11}\equiv\Phi_{02}$ due to $S = S^a_{\;a} =
-2 S_{ab}n^ak^b+S_{ab}m^am^b = 0$ (this point is missing in \cite{AN95}).
When $n^a$ and $k^a$ are swapped, the index 1 in $\Phi_{AB}$ remains unchanged,
while the rest flip $0 \leftrightarrow 2$. It should be mentioned
that we follow Penrose-Rindler~\cite{PR84a} and use the analogous
identification between $\Phi_{AB}$ and $\Phi_{ABCD}$, which is 
different from the choice used in~\cite{AN95}.

Concluding, we present the Lorentz transformations performed in the null
frame $\{k^a, m^a, n^a\}$ are given by \cite{HMP87,HC99}  boosts
\begin{equation}
\tilde{n}^a = \sqrt{A}\,n^a,\;\; \tilde{k}^a =\frac{1}{\sqrt{A}}\,k^a,\;\; \tilde{m}^a = m^a, 
\label{boost}
\end{equation}
where $A>0$, null rotations which leave $n^a$ invariant
\begin{equation}
\tilde{k}^a = k^a + B\,m^a +\frac{1}{2}B^2n^a,\;\; \tilde{m}^a = m^a + B\,n^a, 
\label{nulrot}
\end{equation}
and null rotations which leave $k^a$ invariant, given by
eqs.\,(\ref{nulrot}) with $n^a$ in place of $k^a$ and conversely 
and a new parameter $C$ replacing $B$, where $B,C\in R$. 
These transformations leave the components eqs.\,(\ref{stmet})
of the metric invariant. The Lorentz group $SO(2,1)$ has
three parameters.

In the next section, as required by the Kalhede classification,
we review the algebraic classification
of the Ricci and Cotton-York tensors in terms of Segre types,
and obtain their canonical forms and the correspondent isotropy
groups using the two-component real spinor formalism.

\section{Algebraic Classification}

The algebraic classification of the Ricci tensor 
in three-dimensional manifolds can be obtained by using 
a null triad and the freedom of Lorentz transformations
in order to simplifying as much as possible their
non-vanishing components~\cite{HMP87,HC99}. Thus, 
through Lorentz transformations it is possible to choose a null frame 
$\{k^a, m^a, n^a\}$, where $k^an_a=-1$ and
$m^am_a=1$, such that the Ricci tensor takes one of the following 
canonical forms:
\begin{eqnarray}
\mbox{Segre type}      &        & \mbox{Canonical form} \nonumber \\
\left[11,1\right]      & R_{ab} & = -2\,\alpha\, k_{(a}n_{b)} -\beta(k_ak_b + n_an_b) \nonumber \\
                       &        & + \,\gamma\, m_am_b \label{seg1} \\ 
\left[(11),1\right]    & R_{ab} & = \gamma\, g_{ab}   \nonumber \\
                       &        & +(\gamma-\alpha)\,(k_a + n_a)(k_b + n_b) \label{seg1r} \\ 
\left[1(1,1)\right]    & R_{ab} & = \alpha\, g_{ab} +(\gamma-\alpha)\,m_am_b \label{seg1b} \\
\left[(11,1)\right]    & R_{ab} & = \alpha\,g_{ab} \label{seg1lor} \\
\left[1z\bar{z}\right] & R_{ab} & = -2\alpha\, k_{(a}n_{b)} -\beta(k_ak_b - n_an_b) \nonumber \\
                       &        & + \,\gamma\, m_am_b \label{segz} \\
\left[12\right]        & R_{ab} & = -2\alpha\, k_{(a}n_{b)} +\lambda k_ak_b \nonumber \\ 
                       &        & + \,\gamma\, m_am_b \label{seg2} \\
\left[(12)\right]      & R_{ab} & = \alpha\, g_{ab} +\lambda k_ak_b  \label{seg2nk} \\                       
\left[3\right]         & R_{ab} & = \alpha\, g_{ab} +\mu(k_am_b + m_ak_b)  \label{seg3}
\end{eqnarray}
where $\alpha,\beta,\gamma\in R$ and $\beta\neq 0$ in eq.\,(\ref{segz}). 
It is possible to choose $\lambda=\pm 1$ and $\mu=\pm 1$.
Note that the curvature scalar 
is given by $R=2\,\alpha+\gamma$, except for Segre 
types [(11,1)], [(12)] and [3] where $R =3\,\alpha$.

The above canonical forms correspond to the algebraic classification of $R_{ab}$ 
in terms of Segre types $[1z\bar{z}]$, [12], [3] and [11,1], obtained through
the solution of the eigenvalue problem 
\begin{equation}
(R^a_{\,b} -\lambda g^a_{\,b})v^b = 0, 
\end{equation}
where $\lambda\in C$. 

Therefore, the canonical forms of the Ricci tensor determine
null frames $\{m^a, n^a, k^a\}$, whose vectors are aligned 
with the principal directions of the Ricci tensor. These
are the canonical frames of Karlhede procedure. They are
not uniquely determined, since the Segre types and their
degeneracies have isotropy groups whose elements are the
Lorentz transformations which leave invariant the canonical forms.

By using the quantities defined by eqs.(\ref{phi0})--(\ref{phi2}) and
the canonical forms given by eqs.(\ref{seg1})--(\ref{seg3}), we can obtain
the following canonical forms and isotropy groups of the 
Segre types of the Ricci spinor, given in terms of the non
vanishing $\Phi_{AB}$:
\begin{eqnarray}
\mbox{Segre type}    & \mbox{Canonical form}                & \mbox{Isotropy group} \nonumber \\
\left[11,1\right]    & \Phi_{00} = \Phi_{22},\, \Phi_{11}     & none  \\
\left[(11),1\right]  & \Phi_{00} = \Phi_{22} = 3\,\Phi_{11} & SO(2)  \\
\left[1(1,1)\right]  & \Phi_{11}                            & SO(1,1)  \\
\left[(11,1)\right]  & \Phi_{AB} = 0                        & SO(2,1)  \\ 
\left[1z \bar{z}\right] & \Phi_{00} = -\Phi_{22},\, \Phi_{11} & none \\ 
\left[12\right]      & \Phi_{22} = 1, \Phi_{11}             & none \\
\left[(12)\right]    & \Phi_{22} = 1                        & n.rot.(k_a inv)  \\
\left[3\right]       & \Phi_{12} = 1                        & none
\end{eqnarray}
For Segre types [11,1] and [1z $\bar{z}$] the quantity $\Phi_{11}$ can be zero.

The Segre type $[1z\bar{z}]$, which is the only case to admit
non-real eigenvalues, has three eigenvectors, two complex conjugate
$k^a\pm i n^a$ and one space-like $m^a$, with eigenvalues 
$\alpha\pm i\beta$ ($\beta\neq 0$) and $\gamma$, respectively. The Segre 
type [12] has two eigenvectors, a null $k^a$ and a space-like $m^a$, 
with eigenvalues $\alpha$ and $\gamma$, respectively. The Segre 
type [3] has one null eigenvector $k^a$ with eigenvalue 
$\alpha$~\cite{HMP87,HC99}.

For Segre type [11, 1] the eigenvectors forms a Lorentz
frame $t^a = (k^a + n^a)/\sqrt{2}$, $z^a = (k^a - n^a)/\sqrt{2}$ and $x^a =
m^a$, whose eigenvalues are $\delta=\alpha+\beta$, $\rho=\alpha-\beta$ and $\gamma$,
respectively. This is the only Segre type with time-like
eigenvector. Therefore, it is possible to choose a Lorentz
frame $\{t^a, x^a, z^a\}$ where the Segre type [11, 1] is given by
the alternate form~\cite{HMP87,HC99}
\begin{equation}
R_{ab} = -\delta t_a t_b + \rho z_a z_b + \gamma x_a x_b.
\end{equation} 

The conformal properties of three-dimensional spacetime
are described by the Cotton-York tensor~\cite{GHHM04}
\begin{equation}
C_{ab} =\sqrt{-g}\,\varepsilon_{bcd}\nabla^c (R^d_{\ a} -\frac{1}{4}g^d_{\ a}R),
\end{equation}
where $g =det(g_{ab})$ and $\varepsilon_{abc}$ is the Levi-Civita symbol 
with $\varepsilon_{012} = 1$. It is invariant under conformal transformations of
the metric and vanishes for conformally flat spacetimes. It also satisfies the 
following conditions
\begin{equation}
C_{ab} = C_{ba}, \;\; C^a_{\, a}= 0, \;\; \nabla^b C_{ab} = 0
\end{equation}
of a symmetric, traceless, covariantly conserved tensor.

Since the Cotton-York tensor $C_{ab}$ is symmetric and traceless,
we find that the Cotton-York spinor $\Psi_{ABCD}$ has the same symmetries
as the Ricci spinor $\Phi_{ABCD}$. Thus, it is classified
through the same Segre types, with the same canonical
forms given in terms of the quantities $\Psi_{AB}$, 
which are defined exactly as $\Phi_{AB}$, i.e.,
\begin{eqnarray}
\Psi_{00} &:=& \Psi_{0000} = C_{ab}k^ak^b, \\
\Psi_{22} &:=& \Psi_{1111} = C_{ab}n^an^b, \\
\Psi_{10} &:=& \Psi_{1000} =\frac{1}{\sqrt{2}}C_{ab}m^ak^b, \\
\Psi_{12} &:=& \Psi_{1011} =\frac{1}{\sqrt{2}}C_{ab}m^an^b, \\
\Psi_{11} &:=& \Psi_{0011} =C_{ab}k^an^b. \\
\end{eqnarray}

Note that not only the Segre type, but also the
principal spinors of $\Psi_{ABCD}$, may be different from those
of $\Phi_{ABCD}$.

At last we obtain a final result required in order to
implement the Karlhede procedure using the most efficient
way. It is not necessary to calculate all Cartan's invariants, since 
they are interrelated by both Bianchi and Ricci identities 
and their differential concomitants.
In the next section, we use the spinor formalism to tackle 
the problem of specifying a minimal set of quantities to 
be computed at each step of differentiation of the Karlhede algorithm.

\section{A Complete Minimal set of n-th Curvature Derivatives}

The first step of the Karlhede algorithm involve 
the computation of the covariant derivatives of the Riemann 
curvature. 
For economy in the computations, it is useful to 
specify a minimal set of quantities to be computed at 
each step of differentiation and we will discuss this
problem here. 

Thus, in this section 
Thus we wish to specify a minimal set of components of the 
spinor $n$-th derivatives of the Riemannian curvature such 
that all derivatives of a given order $m$ can be expressed 
algebraically (using sums, products and contractions) 
in terms of these sets for $n\leq m$.

A relevant point to be taken into account when one
needs to compute covariant derivatives of the curvature
tensor is that they are interrelated by both Bianchi
and Ricci identities and their differential concomitants.

The new result given here for three-dimensional spacetimes is the 
analogous to the result obtained by MacCallum and 
{\AA}man~\cite{MA86} for four-dimensional spacetimes. Their result is 
given in terms of two-component complex spinors and includes 
the Weyl spinor and its covariant derivatives. Here we use
two-component real spinors but follows the same approach. 

Although other minimal sets exist (and are related to our 
choice by the application of the Ricci and Bianchi identities) 
the chosen set has two nice properties: it is recursively defined 
(which avoids any need to compute additional $n$-th derivative
quantities in order to find the higher derivatives) and contains 
only totally symmetric spinors (which simplifies storage and 
retrieval algorithms).

Before stating and proving the new result we review 
the derivation of the number of algebraically independent 
quantities that must be given to specify the invariants 
formed from $n$-th derivatives of the Riemannian 
curvature of a general three-dimensional spacetime (which is what 
the spinor components in a canonical frame are). This number is
given by 
\begin{equation}
\frac{3}{2}(n+4)(n+1).
\label{nid}
\end{equation}
The above result is most easily seen by 
considering invariants formed from the coordinate components 
of the $(n+2)$-th derivatives of the metric. 
Since partial derivatives commute, all such derivatives are
expressible in terms of 
$g_{(ab),(c_1,c_2,\ldots,c_{(n+2)})}$, 
where round brackets denote symmetrisation. The number 
of these quantities can easily be computed by considering the 
partition of the indices in each symmetrisation between the 
three different possible values, giving 
\begin{equation}
\frac{6(n+4)!}{2!(n+2)!} = 3(n+4)(n+3).
\label{nimd}
\end{equation}
However, there are still the possible coordinate transformations 
to consider, specified by three functions whose $(n+3)$-th 
derivatives contribute to the $(n+2)$-th derivative of the metric. 
By a similar argument, there are 
\begin{equation}
\frac{3(n+5)!}{2!(n+3)!}=\frac{3}{2}(n+5)(n+4)
\label{nicd} 
\end{equation}
distinct contributions arising in this way. The number 
of independent quantities stated above is the difference 
of eq.\,(\ref{nicd}) and eq.\,(\ref{nimd}). 
 
The number of independent quantities is considerably less than 
the total number of components of the $n$-th derivative which 
is $6\times 3^n$. One can easily compute the cumulative total 
number of these independent quantities up to the $n$-th derivative, which is
\begin{equation}
\frac{1}{2}(n+6)(n+2)(n+1).
\label{indq}
\end{equation}
For $n=6$ (the bound on $p$ given by Cartan for $d=3$) the cumulative total
number of all components of the derivatives is 4374, whereas the number just
derived is 336. Similarly, for $n=5$ (the upper bound on $p$ for metrics of 
Segre types [(11),1], [1(1,1)] and [(12)]) the numbers are 1458 and 231. 
Therefore, it is necessary a computer algebra implementation to carry 
out the calculations.

To end this part, we quote the spinor forms of the Bianchi identities
\begin{equation}
\nabla^{AB}\Phi_{ABCD}+\frac{1}{12}\nabla_{CD}\Lambda = 0
\label{1.3}
\end{equation}
and the Ricci identities
\begin{equation}
\nabla^{N}_{\ (A}\nabla_{B)N}\psi_C = \Phi_{ABCD}\psi^D
+ \frac{\Lambda}{6}(\epsilon_{CA}\psi_B + \epsilon_{CB}\psi_A ).
\label{1.4}
\end{equation}

Let us define the set of $n$-th derivatives $\nabla^nR$ 
to contain the following.
\begin{enumerate}
\item[(i)] 
The totally symmetrised $n$-th covariant derivatives of 
the Ricci spinor:\\
$\nabla_{(AX}\nabla_{BW}\ldots\nabla_{GZ}\Phi_{HKLM)}$.
\item[(ii)]
The totally symmetrised $n$-th covariant derivatives of 
the curvature scalar:\\
$\nabla_{(AX}\nabla_{BW}\ldots\nabla_{GZ)}\Lambda$.
\item[(iii)]
For $n\geq 1$ the totally symmetrised $(n-1)$-th covariant derivative of 
the Cotton-York spinor:\\
$\nabla_{(AX}\nabla_{BW}\ldots\nabla_{GZ}\Psi_{HKLM)}$.
\item[(iv)]
For $n\geq 2$, the d'Alembertian of all quantities in
$\nabla^{(n-2)} R$, i.e.,\\
$\Box\,Q \equiv \nabla^{NN}\nabla_{NN}\,Q$,\\
where $Q$ is a member of $\nabla^{(n-2)} R$.
\end{enumerate}

The new result for three-dimensional spacetimes is obtained  by following 
the same reasoning used by MacCallum and {\AA}man~\cite{MA86}, 
since we have similar relations between the quantities. 
Thus we have the following:

{\em Theorem}. All $n$-th derivatives of the Riemann tensor can 
be expressed algebraically in terms of the elements of $\nabla^rR$ 
for $0\leq r \leq n$ and this is a minimal such set of derivatives.

{\em Proof}. For $n=0$ the statement is merely the decomposition of
the Riemannian curvature spinor and, for $n\geq 1$ the $n$-th derivatives
of the curvature spinor are given by the $n$-th derivatives of the spinors
$\Phi$ and $\Lambda$.

Following MacCallum and {\AA}man, we use the notation $\sim$ for 
the equivalence relation that $n$-th derivatives are equal modulo 
algebraic expressions in the derivatives up to order $(n-1)$.
The Ricci identity shows (see previous section) that for any 
spinor Q, the skew derivative
\begin{eqnarray}
(\nabla^X_{\ A}\nabla^Y_{\ B}-\nabla^Y_{\ B}\nabla^X_{\ A})Q & = & 
\epsilon_{AB}\nabla^{(X}_{\ \ C}\nabla^{Y)C}Q \nonumber \\ 
  && + \epsilon^{XY}\nabla_{Z(A}\nabla^Z_{\ B)}Q \nonumber \\
  &\sim& 0
\label{2.1}
\end{eqnarray}
Consequently all $n$-th derivatives with the same indices on their 
differentiation operators, regardless of the order of these 
operators, are equivalent (under $\sim$). Moreover, a useful
consequence of eq.\,(\ref{2.1}) is
\begin{equation}
\nabla^X_{\ C}\nabla^{YC}Q \sim \epsilon^{XY}\nabla^{ZC}\nabla_{ZC}Q
=\epsilon^{XY}\Box Q
\label{2.2}
\end{equation}
The $n$-th derivatives can be decomposed into their totally 
symmetrised parts and products of the $\epsilon$ spinor with 
$n$-th derivatives having fewer but still symmetrised free indices. 
The totally symmetrised parts are covered by (i)-(ii) above, 
so we need consider only the parts involving contractions. 
We will use induction to prove that all such components are 
algebraically expressible using quantities of the forms (i)-(iv) 
above; the induction hypothesis is in effect used whenever we 
quote eq.\,(\ref{2.1}) or eq.\,(\ref{2.2}), since the equivalence uses terms 
from lower derivatives and we are assuming that these can be 
expressed in terms of quantities of the forms (i)-(iv) above.

The parts involving contractions are symmetric sums of terms, 
each of which involves a contraction. If the contraction indices 
belong to a pair of differentiation operators whose other indices 
are not contracted together, the term can be ignored as a result of 
eq.\,(\ref{2.1}) or is converted by eq.\,(\ref{2.2}) into a term involving contractions 
of both pairs of indices on a certain pair of differentiation operators 
(which will thus form a d'Alembertian). For this second type of 
contraction term, we can bring the d'Alembertian to the left 
(by eq.\,(\ref{2.1})); such terms will then be included in (iv) above. To 
complete the proof that (i)-(iv) are sufficient to represent all 
components of the $n$-th derivative of the Riemann curvature, we 
still have to prove that terms in which the only contractions are 
between differentiation operators and Riemann spinor indices are 
covered by (i)-(iii). We now consider this case.

Using eq.\,(\ref{2.1}) we bring all those differentiation operators which 
are contracted with the Riemann spinor components to the right. We 
can ignore the Ricci scalar (since it has no indices on which to contract). 
By definition of $\Psi$, contraction of a differentiation operator 
with the Ricci spinor leads only to (derivatives of) $\Psi$ and 
the contraction between the Ricci spinor and a differentiation 
operator has a symmetric part which is $\Psi$ and a skew part which 
reduces, by the Bianchi identity eq.\,(\ref{1.3}), to terms of the form (ii). 
The totally symmetrised derivatives of $\Psi$ are (iii) above, so 
we have only to consider terms in which there is a contraction of 
a differentiation operator with $\Psi_{ABCD}$, that is,
with $\nabla^N_{\ (A}\Phi_{BCD)N}$. 
We have now reduced the sufficiency proof to the consideration 
of (derivatives of) $\nabla^A_{\ F}\nabla^N_{\ A}\Phi_{BCDN}$ and 
$\nabla^F_{\ B}\nabla^N_{\ A}\Phi^B_{CDN}$. The first derivative, 
through the previous remarks about terms with contractions 
on differentiation operators, lead to quantities which are 
already included. The second derivative is equivalent, modulo 
d'Alembertian type terms, to a sum of terms of the form 
\begin{eqnarray}
\nabla^F_{\ B}\nabla^N_{\ A}\Phi^B_{\ CDN} &=&
\nabla^F_{\ A}\nabla^N_{\ B}\Phi^B_{\ CDN} \nonumber \\
&&+\epsilon_{AB}\nabla^{FG}\nabla^N_{\ G}\Phi^B_{\ CDN}
\label{2.3}
\end{eqnarray}
the last expression following by the usual decomposition method. 
The first term on the right of eq.\,(\ref{2.3}) reduces, by the Bianchi 
identity eq.\,(\ref{1.3}), to a quantity of the form (ii) and the second 
reduces by eq.\,(\ref{2.1}) and eq.\,(\ref{2.2}) to a quantity of the form (iv).

Having proved that the set given by (i)-(iv) enables all $n$-th 
derivatives to be expressed, we have to show it is minimal. 
This is done simply by a counting argument. It is fairly easy 
to see, by a similar but simpler argument to that given above for
counting invariants, that the constituent parts of $\nabla^nR$ 
as defined above contain respectively $2(n+1)+4$, $2(n+1)$, 
$[2(n-1)+1]+4$ and, assuming the induction hypothesis, 
$3(n+2)(n-1)/2$ real quantities. The total number of quantities 
in $\nabla^nR$ is therefore $3(n+4)(n+1)/2$ as required.
This completes the proof of the theorem.

It should be mention that the theorem and the numbers of components
to which we refer to above are for the general case.

In the next section, all these results are applied to study
the properties of a three-dimensional spacetime. All the
procedures were implemented using the {\em GrTensorII}
package of {\em Maple} computer algebra system.

\section{Three-dimensional homogeneous G\"odel-type spacetimes}

G\"odel's~\cite{godel} solution of Einstein's field equations is the
first cosmological model with rotating matter and closed
time-like curves. It has shown that General Relativity does
not forbid spacetimes with global causal pathologies.
Nevertheless, it can represent rotating objects with physical
meaning when surrounded by more standard spacetimes \cite{BSM98}. 
It should be mentioned that both time-like
and null geodesics are not closed causal curves. The model is
geodesically complete and has neither singularities nor
horizons \cite{HE73}. Due to its peculiarities, G\"odel-type 
solutions have been studied with interest until nowadays with
a fairly large literature.

Recently it has been shown that the three-dimensional
Einstein-Maxwell theory with a cosmological constant
and a Chern-Simons term have G\"odel-type black holes
and particle solutions~\cite{BBCG06}. Furthermore, it was shown
that a one-parameter family of the three-dimensional G\"odel-type
metrics can be seen as arising from a deformation of
anti-de Sitter metric, involving tilting and squashing of
the lightcones. The anti-de Sitter metric appears as the
boundary between the causal and non-causal models~\cite{RS98}.

Actually, the four-dimensional G\"odel spacetime
metric has a direct product structure 
$ds^2_{\,(4)} = ds^2_{\,(3)}+dz^2$, where the three-dimensional metric 
$ds^2_{\,(3)}$ is a particular case of the G\"odel-type line element, 
defined by
\begin{equation}
ds^2_{\,(3)} = -[dt+H(r)d\phi]^2+D^2(r)d\phi^2+dr^2.
\label{goty}
\end{equation}

All four-dimensional G\"odel-type spacetimes that are
homogeneous in space and time (hereafter ST homogeneous),
are characterized by two parameter $m^2$ and $\omega$, where
\begin{itemize}
\item[(i)] $H = \omega\,r^2$, $D = r$, when $m = 0$;
\item[(ii)] $H = (2\,\omega/\mu^2)[1 - cos(\mu r)]$, $D = (1/\mu)sin(\mu r)$,
when $m^2 = -\mu^2 < 0$;
\item[(iii)] $H = (4\,\omega/m^2)sinh^2(mr/2)$, $D = (1/m)sinh(mr)$,
when $m^2 \geq 0$.
\end{itemize}
The constant $\omega$ is the vorticity of these rotating spacetimes.
The G\"odel spacetime is a particular case of the
last class, for which $m^2 = 2\,\omega^2$, whose energy-momentum
tensor $T_{\mu\nu}$ is given by
\begin{eqnarray}
& T_{\mu \nu}=\rho v_{\mu} v_{\nu}\,, \qquad 
  v^{\alpha}=\delta^{\alpha}_{\ 0}\,,&  \label{gdsrc1} \\
& \kappa \rho = - 2 \Lambda = m^{2} = 2\, \omega^{2}\,, \label{gsol} & 
\end{eqnarray}  
where $\kappa$ and $\Lambda$ are, respectively, the Einstein gravitational
and the
cosmological constants, $\rho$ is the fluid density and $v^{\alpha}$ its
four-velocity. The four-dimensional G\"odel
model has a 5-parameter group of isometries with an 1-
parameter isotropy subgroup.

It should be mentioned that the three-dimensional G\"odel-type
line element eq.\,(\ref{goty}), besides being a solution of the
Einstein-Maxwell-Chern-Symon theory, also satisfies the
three-dimensional Einstein's equations for all values of
($m^2,\omega$) and has 4-parameter group of isometries~\cite{BBCG06}.

The (global) causality breakdown in ST homogeneous
four-dimensional G\"odel-type spacetimes depends on the behavior
of $g_{\phi\phi} = D^2(r)-H^2(r)$, since the circles defined by
$t,r,z = const$ are closed time-like curves when $g_{\phi\phi} < 0$
for a certain range of values of r. The causality features
are the following~\cite{CRT88}: (i) for $m^2 < 0$, there is an
infinite sequence of alternating causal and non causal regions;
(ii) for $0\leq m^2 < 4\,\omega^2$, there is only one non causal
region; (iii) for $m^2\geq 4\,\omega^2$, there is no causality problem.
Among these solutions there is the Rebou\c{c}as-Tiomno solution~\cite{RT83}, 
where $m^2 = 4\,\omega^2$, which is conformally flat, has
a 7-parameter group of isometries and is not stably causal.
All models where $m^2 > 4\,\omega^2$ are stably causal. Thus, the
Rebo\c{c}as-Tiomno model is the boundary between the
causal and non-causal models. 

The problem of ST homogeneity of a four-dimensional 
spacetime endowed with a G\"odel-type metric eq.\,(\ref{goty}) was 
investigated under restrictive assumptions on the form
of the Killing vector fields~\cite{RT83,RayTh80,TRA85}. The
result was a set of necessary and sufficient conditions,
which was rederived without assuming any simplifying
hypothesis~\cite{RA87}, by using a complete and invariant description
of spacetimes obtained through the equivalence
problem techniques.

Although the investigation of the causality features of
three-dimensional G\"odel-type spacetimes with parameters
$m^2$ and $\omega$ can be investigated following the approach
used for the four-dimensional case, the situation concerning
the problem of ST homogeneity in its all generality
is completely different. The necessary and sufficient conditions
for ST homogeneity in 4-dimensions~\cite{RA87} were obtained
by using a minimal set of independent covariant
derivatives of the curvature spinor~\cite{MA86}, calculated in a
fixed canonical frame. Unfortunately this approach can
not be used for three-dimensional manifolds, where the Weyl
tensor vanishes identically~\cite{Eisen}, for the following reasons.
First, it requires a different spinor formalism related to
the Lorentz group $SO(2,1)$ of three-dimensional spacetimes.
Second, since the minimal set~\cite{MA86} used depends on the
Weyl spinor and its covariant derivatives, a new minimal
set of independent covariant derivatives of the curvature
spinor must be find. Finally, since the standard frame
was fixed by aligning the frame vectors with the principal
directions of the Weyl spinor, a different canonical
frame fixed by the principal directions of the Ricci spinor
is required.

We show that the ST homogeneous three-dimensional
G\"odel-type manifolds have a 4-parameter group of isometries
and an 1-parameter subgroup of isotropy. The equivalence
of these spacetimes is discussed and they are
found to be characterized by two essential parameters
$m^2$ and $\omega$: identical pairs ($m^2$, $\omega$) correspond to equivalent
(isometric) manifolds. The algebraic classifications
and canonical forms of both the Ricci spinor $\Phi_{AB}$ and the conformal
Cotton-York spinor $\Psi_{AB}$ are presented, by using the formalism
of two-component real spinors \cite{TC94,AN95,TG03}. For a general
pair ($m^2$, $\omega$), the Cotton-York spinor $\Psi_{AB}$ and the Ricci
spinor $\Phi_{AB}$ are both Segre type [(11),1]. The only exceptions
are the conformally flat spacetimes given by the
anti-de Sitter spacetime ($m^2=4\,\omega^2$), where $\Phi_{AB}=0$
and $\Psi_{AB}= 0$, and the spacetime without rotation
($m^2\neq 0$, $\omega=0$), where $\Psi_{AB}=0$. The group of isometries
is also discussed. Finally, the results obtained are
compared with those for the four-dimensional G\"odel-type
ST homogeneous metrics~\cite{RA87}.

In this section we shall consider a three-dimensional pseudo-Riemannian 
manifold M, endowed with a G\"odel-type metric eq.\,(\ref{goty}).

For arbitrary functions $H(r)$ and $D(r)$, the G\"odel-type
three-dimensional metrics are Segre type [11,1] and the
canonical frame is completely fixed. The canonical frame
is found in the following way. First, we calculate the Ricci
spinor $\Phi_{AB}$ in the null triad
\begin{eqnarray}
\theta^1 &=& \omega^1 , \nonumber \\
\theta^2 &=& (\omega^3 - \omega^2)/\sqrt{2} ,\nonumber \\
\theta^3 &=& (\omega^3 + \omega^2)/\sqrt{2} , 
\end{eqnarray}
where $\omega^A$ is a Lorentz triad ($\eta_{AB} = diag (+1, +1,-1)$)
given by 
\begin{equation}
\omega^1 = dr,\;\; \omega^2 = D(r)d\phi,\;\; \omega^3 = dt + H(r) d\phi. 
\end{equation}
We find the non vanishing components $\Phi_{00}$, $\Phi_{22}$ and $\Phi_{11}$.
In order to obtain the canonical form for the Segre type [11,1], 
where $\Phi_{00} = \Phi_{22}$, we perform a boost with
parameter A(r) whose effect on the null frame above can
be stated as
\begin{equation}
\tilde{\theta}^1= \theta^1, \;\; \tilde{\theta}^2=\sqrt{A(r)}\,\theta^2, \;\;
\tilde{\theta}^3=\frac{\theta^3}{\sqrt{A(r)}}
\end{equation}
The canonical frame for the Segre type [11,1] is obtained
by choosing
\begin{equation}
A(r)^2 = \frac{\frac{D''}{D}-(\frac{H'}{D})^2 - (\frac{H'}{D})'}{\frac{D''}{D}
                -(\frac{H'}{D})^2 + (\frac{H'}{D})' }
\end{equation}
where the prime denotes derivative with respect to r.

Using the canonical frame above one finds the following
non vanishing components of the Cartan's scalars, which
correspond to the first step for q = 0 of our algorithm:
\begin{eqnarray}
\Phi_{00} &=& \Phi_{22} \nonumber \\
          &=& \frac{A(r)}{4}\left[\frac{D''}{D}-\left(\frac{H'}{D}\right)^2
          -\left(\frac{H'}{D}\right)' \right], \label{csg1}\\
\Phi_{11} &=& -\frac{1}{12}\left[ \frac{D''}{D}-\left(\frac{H'}{D}\right)^2 \right],\label{csg2} \\
\Lambda   &=& -2\left[\frac{D''}{D}-\frac{1}{4}\left(\frac{H'}{D}\right)^2 \right].\label{csg3}
\end{eqnarray}
For ST homogeneity one finds, from equation eq.\,(\ref{ddim}),
that we must have $t_p$ = 0, that is, the number of
functionally independent functions of the spacetime
coordinates in the set $I_p$ must be zero. Accordingly, all
the above quantities of the minimal set must be constant.
Thus, from eqs.\,(\ref{csg1})--(\ref{csg3}) one easily concludes that
for a three-dimensional G\"odel-type spacetime metric eq.\,(\ref{goty}) to be
ST homogeneous it is necessary that
\begin{equation}
\frac{H'}{D}= \mbox{const}\equiv 2\,\omega, \;\; 
\frac{D''}{D}=\mbox{const}\equiv m^2 .
\label{hcond}
\end{equation}
We shall now show that the above necessary conditions
are also sufficient for ST homogeneity. Indeed, under the
conditions eqs.\,(\ref{hcond}) we obtain that $A(r) = 1$ and
the non-vanishing Cartan's scalars corresponding to the first step for
$q = 0$ of our algorithm reduce to
\begin{eqnarray}
\Phi_{00} &=& \Phi_{22} = 3\,\Phi_{11} = -\frac{1}{4}(m^2-4\,\omega^2) \label{cs1q0} \\
\Lambda  &=& -2\,(m^2-\omega^2) \label{cs2q0}
\end{eqnarray}
where obtain the canonical form $\Phi_{00}=\Phi_{22}=3\,\Phi_{11}$
for Segre type [(11),1].
Therefore, we group the G\"odel-type three-dimensional spacetimes,
according to the relevant parameters $m^2$ and $\omega$, into three
classes:
\begin{itemize}
\item[(i)] 
$m^2\neq 4\,\omega^2$, where $m^2,\omega\neq 0$;
\item[(ii)]
$m^2 = 4\,\omega^2$, where $\omega\neq 0$;
\item[(iii)]
 $m^2 \neq 0$, $\omega = 0$.
\end{itemize}
Now we proceed by carrying out the next steps of the
procedure for testing equivalence in practice, for each
class of G\"odel-type three-dimensional spacetimes.

For the first class, we have that all metrics are Segre
type [(11),1]. Following the algorithm of the previous section,
one needs to find the isotropy group which leaves the
above Cartan's scalars (canonical forms) invariant. The
curvature scalar $\Lambda$ is invariant under the whole Lorentz
group $SO(2,1)$. Thus, the isotropy subgroup $H_0$ is determined
by the Ricci spinor, which is Segre type [(11),1]
and is invariant under the group $SO(2)$ of spatial rotations. So, 
we obtain that $t_0=0$ and that the residual group $H_0$, which 
leaves the above Cartan's scalars invariant, is one-dimensional.

We proceed by carrying out the next step of our practical
procedure, i.e., by calculating the Cotton-York spinor and the 
totally symmetrised covariant derivatives of the Cartan's scalars 
eqs.\,(\ref{cs1q0})--(\ref{cs2q0}), that is, the first step for $q=1$.
One finds the following non-vanishing quantities:
\begin{eqnarray}
\Psi_{00} = \Psi_{22} = 3\,\Psi_{11} &=& 
-\frac{3}{4}\omega(m^2-4\,\omega^2), \label{cs1q1}
\end{eqnarray}
where the Cotton-York spinor is in the canonical form of
Segre type [(11),1].

As no new functionally independent function arose,
$t_0 = t_1$. Besides, the Cartan's scalars eq.\,(\ref{cs1q1}) are
invariant under the same isotropy group (spatial rotations),
i.e. $H_0 = H_1$, and the algorithm stops. Thus we obtain $t_p=0$ and 1-dimensional
isotropy subgroup $H_p=SO(2)$. From eqs.\,(\ref{gdim})-–(\ref{ddim}) one finds
that the three-dimensional G\"odel-type spacetime have a 
four-dimensional group of isometries with a three-dimensional orbit -- the
necessary conditions eqs.\,(\ref{hcond}) are also sufficient for
ST homogeneity.

For the next class ($m^2 = 4\,\omega^2, \omega\neq 0$), following
the algorithm, we obtain that 
\begin{eqnarray}
\Phi_{AB} = 0, \label{cs1mq0} \\ 
\Lambda = -\frac{3}{2}m^2 \label{cs2mq0}.
\end{eqnarray}
Thus $t_0 = 0$ and $dim(H_0) = 3$, since the group of
invariance of $\Phi_{AB}$ is now the Lorentz group
$SO(2,1)$. Considering that the Cotton-York spinor
$\Psi_{AB} = 0$ and that all derivatives of the Cartan's 
scalars vanishes identically, 
the process terminate. This is the anti-de Sitter
three-dimensional spacetime. It is conformally flat and
has a 6-dimensional isometry group with a three-dimensional
isotropy subgroup and acts on a three-dimensional orbit. Therefore,
according to eq.\,(\ref{ddim}), it is ST homogeneous. Note that the
four-dimensional spacetime with the same values of $m^2,\omega$
is the conformally flat Rebou\c{c}as-Tiomno spacetime 
with a 7-parameter isometry group.

Finally, for the last class ($m^2\neq 0,\omega=0$), that is,
the G\"odel-type spacetimes without rotation, we obtain 
for the first step of the algorithm for q = 0 that
\begin{eqnarray}
\Phi_{00} &=& \Phi_{22} = 3\,\Phi_{11} = -\frac{1}{4}m^2 \label{csf1q0}  \\
\Lambda  &=& -2m^2 \label{csf2q0}
\end{eqnarray}  
where $\Phi_{AB}$ is in canonical form for Segre type [(11),1].
Therefore, we find again that $t_0 = 0$ and that $H_0$ is
the group $SO(2)$ of spatial rotations. 

Following the algorithm, we obtain once more that the Cotton-York
spinor $\Psi_{AB}=0$ and that all derivatives vanishes identically, and 
the process terminate. Since $t_p=0$ and $H_p=SO(2)$, we obtain again
a 4-parameter isometry group with a 1-dimensional isotropy
subgroup. Therefore, it is a ST homogeneous conformally flat 
spacetime. Notice that the four-dimensional G\"odel-type spacetime 
without rotation is not conformally flat and has a 6-dimensional 
isometry group.

The above results can be summarized in the following theorems:
\begin{theorem}  
The necessary and sufficient conditions for
a three-dimensional G\"odel-type spacetime to be ST (locally)
homogeneous are those given by equations eqs.\,(\ref{hcond}).
\end{theorem}
\begin{theorem}
All ST locally homogeneous three-dimensional G\"odel-type spacetimes 
admit a four-dimensional group of isometries with an 1-parameter 
isotropy subgroup and are characterized by two independent parameters $m^2$ and
$\omega$: identical pairs ($m^2,\omega$) specify equivalent (isometric) spacetimes.
\end{theorem}
It is worth emphasizing that eqs.\,(\ref{cs1q0})-–(\ref{csf2q0}) are
related to the corresponding equations for four-dimensional
G\"odel-type spacetimes (eqs.\,(3.12)-–(3.15) and eqs.\,(3.18)-–(3.25) in 
\cite{RA87}). Therefore, our study here can be considered
as the lower dimensional analogous of results in \cite{RA87}. Thus, 
for example, the above theorems 1 and 2 relates to the corresponding 
theorems in \cite{RA87} (theorems 1 and 2 on page 891).

For all three-dimensional ST homogeneous G\"odel-type
spacetimes the Segre type of the Ricci spinor is [(11),1].
The only exception is the anti-de Sitter spacetime, obtained
when $m^2 = 4\,\omega^2$, whose isometry group is 6-dimensional. 
This is the only condition where the isometry
group has dimension higher than four. The Cotton-
York spinor has also the same Segre type [(11),1] for
all G\"odel-type, except when the spacetime is anti-de
Sitter ($m^2 = 4\,\omega^2$) and when the spacetime has no rotation
($\omega = 0, m^2\neq 0$), since both are conformally flat
($\Psi_{AB} = 0$).

For the sake of completeness, we should mention that
the equivalence problem techniques were used to investigate
the five-dimensional G\"odel-type and generalized G\"odel-type
pseudo-Riemannian spacetimes~\cite{Reb1,Reb2}. Furthermore,
G\"odel-type solutions with torsion were also investigated~\cite{AFRM98} 
through the equivalence problem techniques for Riemann-Cartan
spacetimes~\cite{frt,frm,frm1}. 

To conclude, we should like to emphasize that as no
field equations were used to show the above results, they
are valid for every three-dimensional pseudo-Riemannian
G\"odel-type solution regardless of the theory of 
gravitation considered.

\section{Final Remarks}

It should be noted that the equivalence problem techniques,
which we have used in this work, can certainly be used in 
more general contexts. Among possible applications we
mention, especially, that the equivalence problem techniques
for both general relativity and three-dimensional gravitation
can be used to investigate their interconnections in a 
coordinate invariant way. \\

\section*{Acknowledgments}

F. C. Sousa gratefully acknowledge financial assistance form CAPES.
C. Romero would like to thank CNPq for partial financial support.
Thanks also go to the referees for their suggestions and constructive
comments.

\end{document}